\documentclass[aps,prl,preprint,groupedaddress]{revtex4}
\usepackage{graphicx}

\begin{document}

%\preprint{}

\title{The r-Process without Excess Neutrons}


\author{Bradley S. Meyer}
\email[]{brad@photon.phys.clemson.edu}
\homepage[]{http://photon.phys.clemson.edu/wwwpages/meyer.html}
%\thanks{}
\affiliation{Department of Physics and Astronomy, Clemson University,
Clemson, SC 29634-0978}


\date{\today}

\begin{abstract}
The r-process of nucleosynthesis requires a large neutron-to-seed nucleus
abundance ratio.  This does not, however, require that there be
an excess of neutrons over protons.  If the expansion of the
matter is sufficiently rapid and the entropy per nucleon is sufficiently
high, the nucleosynthesis enters a heavy-element synthesis
regime heretofore unexplored.
In this extreme regime, characterized by a persistent disequilibrium between
free nucleons and the abundant alpha particles, heavy r-process nuclei
can form even in matter with more protons than neutrons.  This observation
bears on the issue of the site of the r-process, on the variability of
abundance yields from r-process events, and on constraints on
neutrino physics derived from nucleosynthesis.  It also clarifies the
difference between nucleosynthesis in the early universe and that in
less extreme stellar explosive environments.
\end{abstract}

%\keywords{}

\maketitle

\section{}

Despite years of study, a proper understanding of the provenance of the
solar system's r-process isotopes remains elusive.  These nuclei,
comprising roughly half the isotopes heavier than iron, are
suspected to have formed in environments expanding rapidly from conditions
of high temperature and density.  During this expansion, heavy seed nuclei
form. If excess neutrons remain after seed production, each nucleus is
subsequently able to capture a large number of neutrons, thereby producing
the r-process isotopes.  This basic
mechanism has long been understood
\cite{1957RvMP...29..547B,1957PASP...69..201C}.  The difficulty in
determining the r-process site lies rather in finding a realistic astrophysical
setting that can generate a sufficiently large neutron-to-seed ratio.

The requirement of a large neutron-to-seed ratio for a successful r-process
has long suggested that the r-process environment must have a substantial
excess of neutrons over protons.  The degree of this excess depends on the
entropy per nucleon and the expansion timescale.  For low-entropy (entropy
per nucleon $s/k_B \lesssim 1$) material,
seed nucleus production is highly efficient.  The material
must therefore be quite neutron rich to have a large enough neutron-to-seed
ratio.  Such material typically requires an electron-to-nucleon ratio
$Y_e \approx 0.1-0.2$.  It should be noted that $Y_e$ is, for charge-neutral
matter, the fraction of all nucleons that are protons. 
While ejection of neutronized matter from supernova cores (e.g.,
\cite{1976A&A....52...63H,2001ApJ...562..880S}) or from colliding
neutron stars (e.g., \cite{2001NuPhA.688..344R})
are promising scenarios for such low-entropy r-processes,
current models have not yet established that enough neutron-rich matter is
ejected to explain the solar system's r-process abundances.

For higher entropies ($s/k_B \ge 100$)
and faster expansions, such as are thought to occur in neutrino-heated
ejecta from proto-neutron stars,
seed nucleus production is less efficient, and the r-process may occur for
larger $Y_e$, typically in the range $Y_e \approx 0.35-0.45$ (for example,
see Refs. \cite{1994ApJ...433..229W,1994A&A...286..857T}). 
Recent r-process calculations in the context of general-relativistic wind models
show great promise.  One group obtains realistic r-process yields
for $Y_e = 0.4$ \cite{2001ApJ...554..578W}.  Another group argues that realistic
supernova neutrino spectra
may not allow $Y_e$ to get so low in the winds and that, if the r-process occurs in these
environments, $Y_e$ is more likely in the range 0.47-0.495
\cite{2001ApJ...562..887T}.
A higher-$Y_e$ r-process
has two advantanges.  First, it does not suffer from the overproduction of $N=50$ isotones
common in wind models with lower $Y_e$ (e.g.,
\cite{1994ApJ...433..229W,1994A&A...286..857T}).
Second, it is less susceptible than lower-$Y_e$ r-processes
to the detrimental effects of the neutrino-induced
``alpha effect'' (e.g., \cite{1998PhRvC..58.3696M}),
which may drastically diminish the neutron-richness of
wind matter during seed nucleus production.  The issue for the
higher-$Y_e$ r-process is
whether the winds can attain high enough entropy and fast enough expansion
rates to make heavy r-process isotopes.

The purpose of this paper is to point out that a high neutron-to-seed ratio
for the r-process does not necessarily require that the environment be neutron
rich.  In fact, heavy r-process nuclei can form in
environments with equal numbers of neutrons and protons ($Y_e = 0.5$)
or even with excess protons ($Y_e > 0.5$).  The nucleosynthesis is qualitatively
different from that in previously studied r-process
scenarios because of a persistent disequilibrium between free nucleons and alpha
particles.  This observation has interesting implications for the site of the
r-process, the variability of r-process yields, and constraints on neutrino
physics drawn from nucleosynthesis.  It also clarifies the difference between
explosive nucleosynthesis in the early universe and in stars.

\begin{figure}
\includegraphics{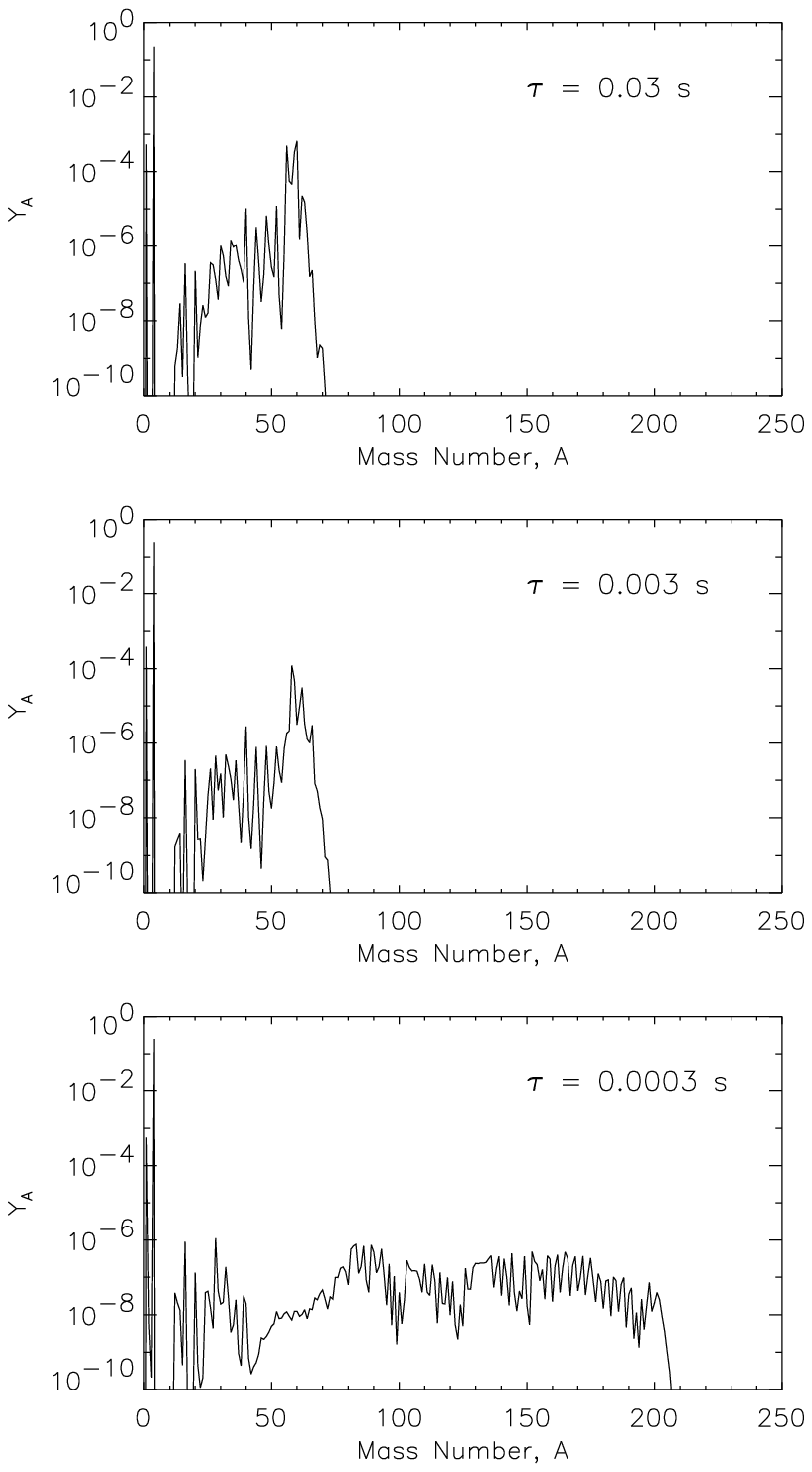}
\caption{\label{fig:abund}Final abundances per nucleon
versus mass number $A$ for increasingly rapid expansions.}
\end{figure}

In order to explore this issue, calculations were made with the Clemson
nucleosynthesis code \cite{1995ApJ...449L..55M}.
This code has been recently updated to
use the NACRE \cite{1999NuPhA.656....3A} and NON-SMOKER
\cite{2000ADNDT..75....1R} rate compilations.
Figure \ref{fig:abund} shows the final abundances per nucleon
(as a function of mass number) for three calculations, each
with $Y_e = 0.5$ and entropy per nucleon $s/k_B = 150$.  In each
calculation, the expanding material was modeled to begin at temperature
above $T_9 = T / 10^9\ {\rm K} = 10$.  The material was taken to expand
with constant entropy and the density to fall with time $t$ as
\begin{equation}
\rho(t) = \rho_1 \exp(-t/\tau) + \rho_2 \left({\Delta \over \Delta
+t}\right)^2
\label{eq:rho}
\end{equation}
where $\rho_1 + \rho_2$ is the density at time $t=0$ and $\Delta$
was chosen so that the two terms on the right-hand side of Eq.
(\ref{eq:rho}) were equal near $T_9 =2$.
This form of expansion models neutrino-driven winds from
neutron stars, which initially expand
roughly exponentially (the first term in Eq.
[\ref{eq:rho}]) but then may evolve to an outflow with constant velocity
(the second term in Eq. [\ref{eq:rho}]).  At each timestep, the temperature
was determined from the composition, entropy, and density
by iteration, as described in Ref.  \cite{1997ApJS..112..199M}.

As Figure \ref{fig:abund} shows, expansions with $\tau = 0.03$ or 0.003
seconds produce significant quantities of $^4$He and iron group nuclei
(particularly isotopes of nickel), as expected for high entropy and
$Y_e = 0.5$ matter.  For the expansion with a $\tau$ of 0.0003 seconds,
however, considerably heavier nuclei form.  While the resulting abundance
pattern does not match the solar system r-process abundance
distribution in detail, many heavy r-process nuclei have been
synthesized.

\begin{figure}
\includegraphics{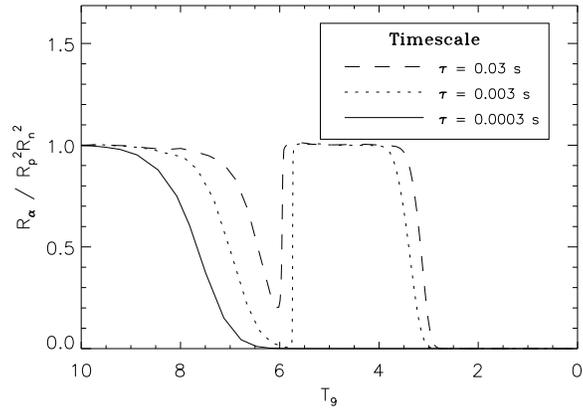}
\caption{\label{fig:rnpa}Equilibrium of neutrons, protons, and alpha
particles during the three expansions of Fig.~\ref{fig:abund}.}
\end{figure}

The nuclear dynamics that allows matter without neutron excess to produce
heavy r-process nuclei is qualitatively different from that in more standard
r-processes.  This is best illustrated in Figure \ref{fig:rnpa}, which
shows the quantity $R_\alpha / R_p^2 R_n^2$ in the three calculations.
The quantity $R_i$ at a particular point in an expansion
is the ratio of the abundance of species $i$ (with proton number $Z_i$ and
neutron number $N_i$) in the
actual reaction network to the abundance that species would have in
nuclear statistical equilibrium at the same temperature, density, and
$Y_e$.  $R_i / R_p^{Z_i} R_n^{N_i}$ is unity if the free nucleons are in
equilibrium with species $i$ and less than unity if species $i$
is underabundant relative to equilibrium
\cite{1998ApJ...498..808M}.  In slower and
lower-entropy expansions, reactions converting nucleons into alpha
particles and back proceed rapidly enough to maintain an equilibrium among
these species down to relatively low temperatures ($T_9 \approx 3-4$).  The
result is that $R_\alpha / R_p^2 R_n^2$ stays near unity throughout much of
the expansion (cf. Fig. 8 of Ref. \cite{1998ApJ...498..808M}).
In the present calculations
this is not the case.  The equilibrium begins to fail as the material cools
below $T_9 \approx 9$.  In the $\tau = 0.03$ and 0.003 s expansions, this
equilibrium recovers around $T_9 = 6$ and then persists down to $T_9
\approx 3.5$.  In the $\tau = 0.0003$ s expansion, the equilibrium never
recovers.

\begin{figure}
\includegraphics{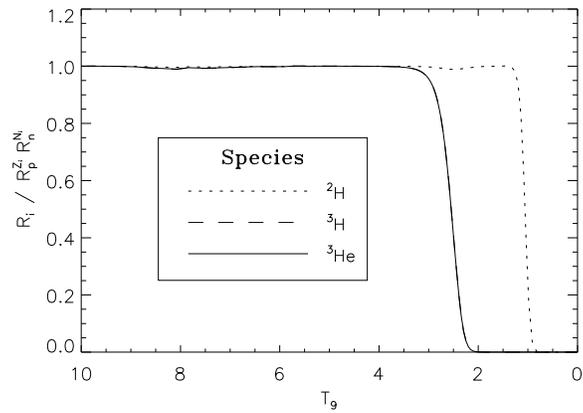}%
\caption{\label{fig:rdt3}The equilibrium among the neutrons, protons,
$^2$H, $^3$H, and $^3$He for $\tau = 0.0003$ s.  The $^3$H
curve lies beneath the $^3$He curve.}
\end{figure}

The equilibrium between the nucleons and alpha particles first fails near
$T_9 \approx 9$ because of the high entropy and low abundance of light
nuclear species.  Figure \ref{fig:rdt3} shows that $^2$H, $^3$H, and $^3$He
all remain in excellent equilibrium with the nucleons in the $\tau =
0.0003$ s expansion down to $T_9 \approx
3$ for the heavier two species
and $T_9 \approx 1$ for deuterium.  These isotopes
behave similarly in the two slower expansions.  Because of the
high entropy, however, the light isotope
abundances are extremely low.  Production of
$^4$He occurs primarily via capture
on these species, so, as the temperature falls,
their low abundances fail to maintain
$^4$He at its equilibrium abundance.

\begin{figure}[top]
\includegraphics{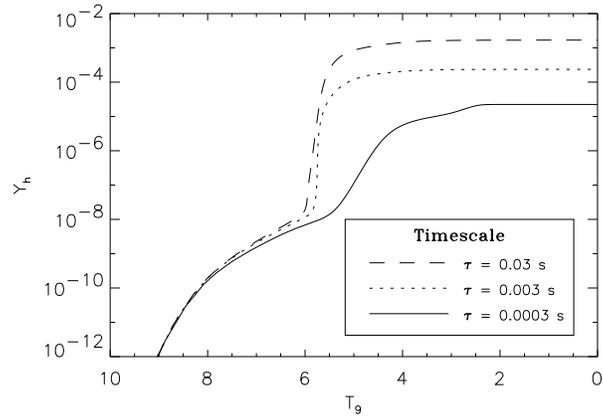}%
\caption{\label{fig:yh}The abundance of heavy nuclei versus temperature
for the three calculations of FIG.~\ref{fig:abund}}
\end{figure}

The return to equilibrium in the two slower expansions happens because of
the production of heavy nuclei.  As Figure \ref{fig:yh} shows, these two
expansions begin producing significant numbers of heavy nuclei near $T_9
\approx 6$.  Once sufficiently many heavy nuclei
have formed, they are able to catalyze the equilibrium between the nucleons
and $^4$He via fast reaction cycles such as
$^{62}{\rm Ni}(p,\gamma)^{63}{\rm Cu}(n,\gamma)^{64}{\rm Cu}(n,\gamma)^{65}{\rm
Cu}(p,\alpha)^{62}{\rm Ni}$ and their inverses.
For the $\tau = 0.0003$ s expansion,
however, considerably fewer nuclei form.  The smaller number of heavy
nuclei are then unable to restore the equilibrium between the nucleons
and $^4$He. 

\begin{figure}
\includegraphics{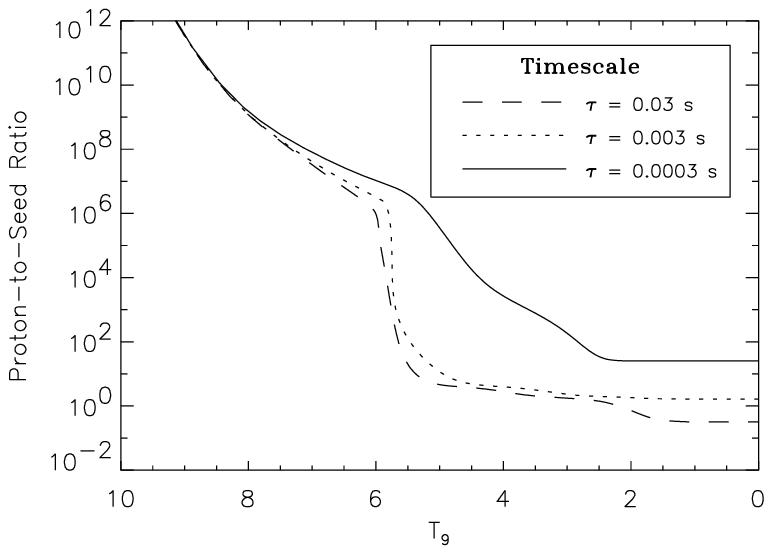}%
\caption{\label{fig:rp} The proton-to-seed ratio versus temperature
for the three calculations of FIG.~\ref{fig:abund}}
\end{figure}

\begin{figure}
\includegraphics{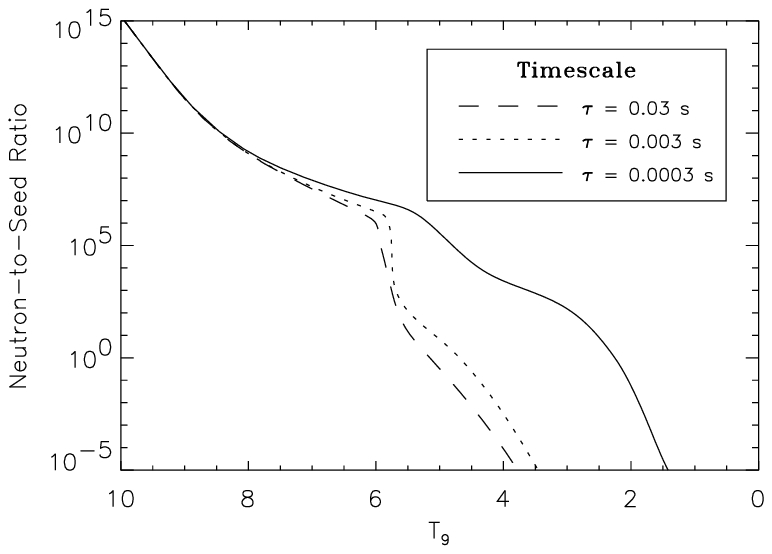}%
\caption{\label{fig:rn} The neutron-to-seed ratio versus temperature
for the three calculations of FIG.~\ref{fig:abund}}
\end{figure}

For temperatures below $T_9 \approx 6$, equilibrium in these expansions
strongly favors locking most free nucleons into $^4$He.  In the $\tau =
0.03$ and $0.003$ s expansions, then, few free nucleons are available at
lower temperatures for subsequent capture.  By contrast, because the
equilibrium fails early for the $\tau = 0.0003$ s expansion and is never
restored, the few free
nuclei that form co-exist with a huge overabundance of free neutrons and
protons down to low temperature.  This is shown in Figures \ref{fig:rp}
and \ref{fig:rn}.
At $T_9 = 4$, there are roughly $10^3$ times more free protons and
$10^6$ times more
neutrons per seed nucleus in the $\tau = 0.0003$ s expansion than in the
two slower ones.  This excess of nucleons leads to heavier seed nuclei than
would be expected in the less extreme
expansions.  Furthermore, the free neutrons are available at lower
temperatures for r-process nucleosynthesis. 
More details of the three expansions, including movies,
are available in the electronic addendum to this paper \cite{epaps}.

\begin{figure}
\includegraphics{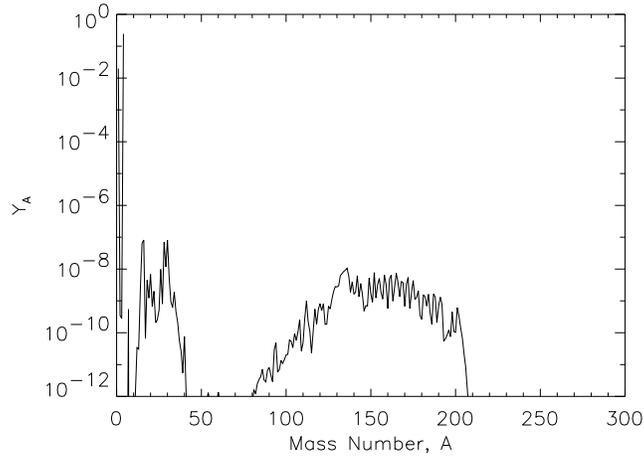}%
\caption{\label{fig:prich} The final abundances per nucleon
as a function of nuclear mass
number for the calculation with $Y_e = 0.505$, $s/k_B = 200$, and
$\tau = 0.0003 s$.
}
\end{figure}

The presence of a persistent disequilibrium between free nucleons and alpha
particles can even allow proton-rich matter to produce heavy r-process nuclei.
For example,
figure \ref{fig:prich} shows the final abundances for a calculation with
$\tau = 0.0003$ s, $s/k_B = 200$, and $Y_e = 0.505$.  The overall yield is
low, and the final distribution does not match the solar r-process
pattern.  Nevertheless, heavy r-process nuclei have formed even though the
matter, in bulk, actually has more protons than neutrons.

Although the expansion timescales and/or entropies
needed for an r-process without excess
neutrons are considerably more extreme than current models yield
(e.g., \cite{2001ApJ...562..887T}),
the present results should nevertheless
encourage further study of supernova models with rapid ejection of
high-entropy but nearly symmetric ($Y_e \approx 0.5$) matter.
Additional nucleosynthesis calculations are also needed to explore
conditions required for a disequilibrium between free
nucleons and alpha particles and the resulting r-process
abundances.  The presence or absence of such a disequilibrium can cause the
final r-process abundances to be quite sensitive to the expansion parameters,
especially for slightly neutron-rich matter.
For example, an expansion with $s/k_B = 150$, $\tau=0.0007$ s,
and $Y_e=0.4975$ predominantly forms third-peak nuclei
(mass number $A \approx 195$)
because of the persistent disequilibrium between nucleons and alpha particles
while a calculation with $s/k_B = 150$, $\tau=0.0008$ s,
and $Y_e=0.4975$ makes a strong second r-process
peak ($A \approx 130$) but few heavier nuclei
because the nucleon-alpha particle
equilibrium is restored \cite{epaps}.  Such sensitivity could result in
variability of r-process yields from nucleosynthesis event to
event and provide a natural explanation for the two r-process component
scenario often invoked to explain the abundances of
extinct short-lived r-process
radioactivities in meteorites (e.g., \cite{2000PhR...333...77Q}).
It should also be clear that since heavy r-process nuclei can form even
in matter with
$Y_e > 0.5$, constraints on the properties of neutrinos, as inferred from
r-process yields (e.g., \cite{1993PhRvL..71.1965Q}), must be made with caution.

The above considerations clarify the fundamental difference between
``normal'' explosive nucleosynthesis in stars and nucleosynthesis in the
early universe.  In ``normal''
explosive stellar nucleosynthesis, conditions are not too extreme,
and the free nucleons
typically remain in equilibrium with the $^4$He nuclei to relatively low
temperature.  This leaves few free protons or neutrons available for subsequent
modification of the abundances, unless the material is significantly neutron
or proton rich.  By contrast, in the early universe, $^4$He would dominate
the equilibrium abundances at $T_9 \approx 3$ and lock up all free
neutrons, resulting in a $^4$He mass fraction greater than 30\% \cite{epaps}.
The extremely high entropy, however, keeps the abundance of
$^2$H, $^3$H, and $^3$He low, thereby preventing $^4$He from attaining its
high equilibrium value.  Synthesis of $^4$He occurs only several minutes
later when the temperature drops below $T_9 \approx 1$.
At this point, the equilibrium between the nucleons and $^2$H breaks and
non-equilibrium nuclear flows build up $^4$He.  Because the neutrons remain
free due to the lack of equilibrium between free nucleons and $^4$He, their
abundance declines by beta decay over the several minutes period from $T_9 =3$
to $T_9 \approx 1$, thereby leading to a $^4$He mass fraction of
approximately 25\%. 

Further modeling will establish whether the solar system's r-process
isotopes formed in a disequilibrium between the free
nucleons and alpha particles.  Regardless of the outcome of
such modeling, however,
a certain irony attends the
fact that the synthesis of heavy r-process nuclei
can be in its fundamental aspect
more similar to the formation of nature's lightest isotopes in the early
universe than to production of heavier isotopes such as $^{56}$Fe
in less extreme explosive processes.

\begin{acknowledgments}
The author gratefully acknowledges conversations with A. Burrows and T.
Thompson and assistance from
 L.-S. The and S. Chellapilla.
This research was
supported by NASA grant NAG5-10454, NSF grant AST 98-19877,
and a SciDAC grant from the High Energy Nuclear Physics Division of the DOE.
The author also thanks the
Institute for Nuclear Theory at the University of Washington for its
hospitality and the Department of Energy for partial support during the
completion of this work.
\end{acknowledgments}


% Surround figure environment with turnpage environment for landscape
% figure
%\begin{turnpage}
%\end{turnpage}

% Surround figure environment with turnpage environment for landscape
% figure
%\begin{turnpage}
%\end{turnpage}

% Surround figure environment with turnpage environment for landscape
% figure
%\begin{turnpage}
%\end{turnpage}

% Surround figure environment with turnpage environment for landscape
% figure
%\begin{turnpage}
%\end{turnpage}

% Surround figure environment with turnpage environment for landscape
% figure
%\begin{turnpage}
%\end{turnpage}

%
% ****** End of file template.aps ******

\end{document}